\documentclass[fleqn,usenatbib]{mnras}

\usepackage{newtxtext,newtxmath}

\usepackage[T1]{fontenc}
\usepackage{ae,aecompl}

\usepackage{graphicx}	
\usepackage{amsmath}	
\usepackage{amssymb}	
\newcommand{\myclus}{RXCJ0232.2-4420\,\,}

\title[Transition mini-halo in \myclus]{A radio halo surrounding the Brightest Cluster Galaxy in \myclus: { a mini-halo in transition ?}}

\author[Kale et al.]{Ruta Kale,$^{1}$\thanks{E-mail: ruta@ncra.tifr.res.in}
Krishna M. Shende,$^{1}$
and
Viral Parekh$^{2}$
\\
$^{1}$National Centre for Radio Astrophysics, Tata Institute of Fundamental Research, S. P. Pune University Campus, Ganeshkhind, \\Pune 411007, India\\
$^{2}$Department of Physics and Electronics, Rhodes University, PO Box 94, Grahamstown, 6140, South Africa\\
}

\date{Accepted XXX. Received YYY; in original form ZZZ}

\pubyear{2019}
\begin{document}
\label{firstpage}
\pagerange{\pageref{firstpage}--\pageref{lastpage}}
\maketitle

\begin{abstract}
Diffuse radio sources associated with the intra-cluster medium are direct probes of 
    the cosmic ray electrons and magnetic fields. 
We report the discovery of a diffuse radio source in the galaxy cluster RXCJ0232.2-4420 
(SPT-CL J0232-4421, $z=0.2836$) using 606 MHz observations with the Giant Metrewave Radio Telescope.
The diffuse radio source surrounds the Brightest Cluster Galaxy in the cluster like typical 
radio mini-halos. However the total extent of it is $550\times800$ kpc$^{2}$, which is larger than mini-halos and {similar to that} of radio halos. 
The BCG itself is also a radio source with a marginally resolved core at $7''$ (30 kpc) resolution. 
We measure the 606 MHz flux density of the RH to be $52\pm5$ mJy. Assuming a spectral index of 1.3, the 
1.4 GHz radio power is  $4.5 \times 10^{24}$ W Hz$^{-1}$.
The dynamical state of the cluster has been inferred to be "relaxed" and also as "complex" 
depending on the classification methods based on the morphology of the X-ray surface brightness.
This system thus seems to be in the transition phase from a mini-halo to a radio halo.
\end{abstract}




\section{Introduction}
The origin and evolution of cosmic rays and magnetic fields in 
galaxy clusters is a long standing puzzle. Diffuse radio emission 
on cluster-wide scales (megaparsec) associated with the intra-cluster medium (ICM) 
and not with individual galaxies are direct probes of these components. 
These are typically classified into three classes: radio halos, radio relics and 
mini-halos \citep{bru14}. Radio relics are believed to be direct tracers of cluster merger shocks 
and observational evidence \citep[e. g.][]{wee10,kal12,str14} strongly supports this idea though the details of the acceleration mechanism are still under debate \citep[e. g.][]{bru14,guo14}.

Radio halos and mini-halos nearly circular in morphology and located co-spatially with the 
X-ray emission from the ICM \citep[e. g.][]{fer12,bru14}. Mini-halos are 100 - 300 kpc in size and are 
found in cool-core clusters \citep[e. g.][]{gia17}. They are always found to surround the Brightest 
Cluster Galaxy (BCG). Radio halos are much more extended and can reach sizes of 1- 2 Mpc 
\citep{bru14}. These have been found with higher probability in massive and merging clusters 
\citep{cas13}.

A component of the relativistic electron population that powers mini-halos and radio halos 
comes from the hadronic collisions in the ICM \citep[e. g.][]{den80,bru01,git04,kes10}. 
For radio halos it has been found that this component alone is not sufficient and an additional 
form of turbulent re-acceleration is required \citep{don10,bru12}. 
Cluster mergers are a natural origin for the 
turbulence involved and thus the turbulent re-acceleration model can explain the high occurrence of radio 
halos in massive and merging clusters \citep[e. g.][]{schli87,bru01,cas13}. The mini-halos on the other hand 
could either be exclusively of hadronic origin or may need turbulence driven by sloshing of sub-clusters or minor mergers 
\citep{git04,zuh14,maz08,gia14,jacob2017,gia17}. 

Although the radio halos and mini-halos are known to occur in merging and relaxed clusters, respectively, 
there are systems which are intermediate in their merging status. It has been discussed that a transition from a mini-halo to a radio halo triggered by a merger is a possibility \citep{bru14,wee19}. The transition can also mark the change from a hadronically dominated source to that dominated by turbulent re-acceleration.

In this letter we report the detection of a diffuse radio source which has the properties of mini-halos 
and also of radio halos that is hosted by the cluster \myclus (Table~\ref{clusprop}).
This cluster was discovered in the ROSAT all sky survey \citep{Cruddace2002} and is known 
to contain two brightest cluster galaxies (BCGs) separated by 100 kpc \citep{2008A&A...483..727P}. 
It is part of the Archive of Chandra Cluster Entropy Profile Tables (ACCEPT) sample \citep{cav09} and they report a central entropy ($K_0$) of $44.62 \pm 12.42$ keV cm$^{2}$ for this cluster. 

We assume a $\Lambda$CDM cosmology with $\mathrm{H}_{0} = 71$ km s$^{-1}$ Mpc$^{-1}$, 
$\Omega_\mathrm{M} = 0.27$ and $\Omega_\Lambda =0.73$. This implies a scale of $4.24$ kpc arcsec$^{-1}$ 
at the redshift 0.2836 of \myclus. The luminosity distance is $D_{\rm L}= 1444$ Mpc.

\begin{table}
	\centering
	\caption{Properties of \myclus. Notes: $^{\dag}$ \citet[][and references therein.]{Bleem2015}
$^{*}$\citet{Lagana2013}}
	\label{clusprop}
	\begin{tabular}{lccrc} 
		\hline
        RA(J2000)(h m s) & 02 32 18.7 \\
        DEC(J2000) ($^\circ$ $^{\prime}$ $^{\prime\prime}$) & -44 20 41 \\
        Redshift ($z$) & 0.2836$^{\dag}$ \\
        $L_{X[0.1-2.4 \mathrm{keV}]}$ (erg s$^{-1}$) &$13.3\times10^{44} $ \\
        M$_{500}(10^{14}$ M$_\odot$) & $12.01\pm1.80^{\dag}$ \\
        kT(keV) & $8\pm1.4^{*}$\\
        \hline
        Radio halo &\\ 
        Size & $550 \times 800$ kpc$^{2}$\\
            $S_{606\mathrm{MHz}}$ & $52 \pm 5$ mJy\\
            $P_{1.4\mathrm{GHz}}$ &  $4.5 \times 10^{24}$ W Hz$^{-1}$\\
	\hline
	\end{tabular}
\end{table}

\section{Observations and data reduction}
We have observed cluster \myclus using the GMRT at 610 MHz with 33 MHz bandwidth under observation 
cycle 31 on 18-Feb-2017 ($31\_065$ PI R. Kale). The cluster was observed for 6 hours with the 
integration time of 16.10 s. The data were recorded in 256 frequency channels.
We used NRAO's Common Astronomy Software Applications (CASA) package for data analysis. 
A semi-automated pipeline was written for the analysis in python that utilised CASA tasks.
The standard steps of flagging (removal of bad data), calibration of complex gains and bandpass 
were carried out. The flux scale of Perley-Butler 2017 was used for absolute flux calibration.
Several rounds of phase only self-calibration followed by amplitude and phase self-calibration were 
carried out to improve the sensitivity. We imaged the final visibilities with the weighting scheme 
"briggs" \citep{Briggs1995} with robust $=0$. We imaged the discrete sources using the 
uv-distance cut of $>5$ k$\lambda$ and subtracted them from the visibilities. A low resolution 
image of the diffuse source using the uv-distance cut of $<10$ k$\lambda$ was produced. 
It was convolved to a circular beam of $20''\times20''$ for further analysis. The primary beam 
correction was applied to all the images that were used in the analysis.

\section{Results}
\begin{figure}
    \centering
                \includegraphics[width= 8 cm]{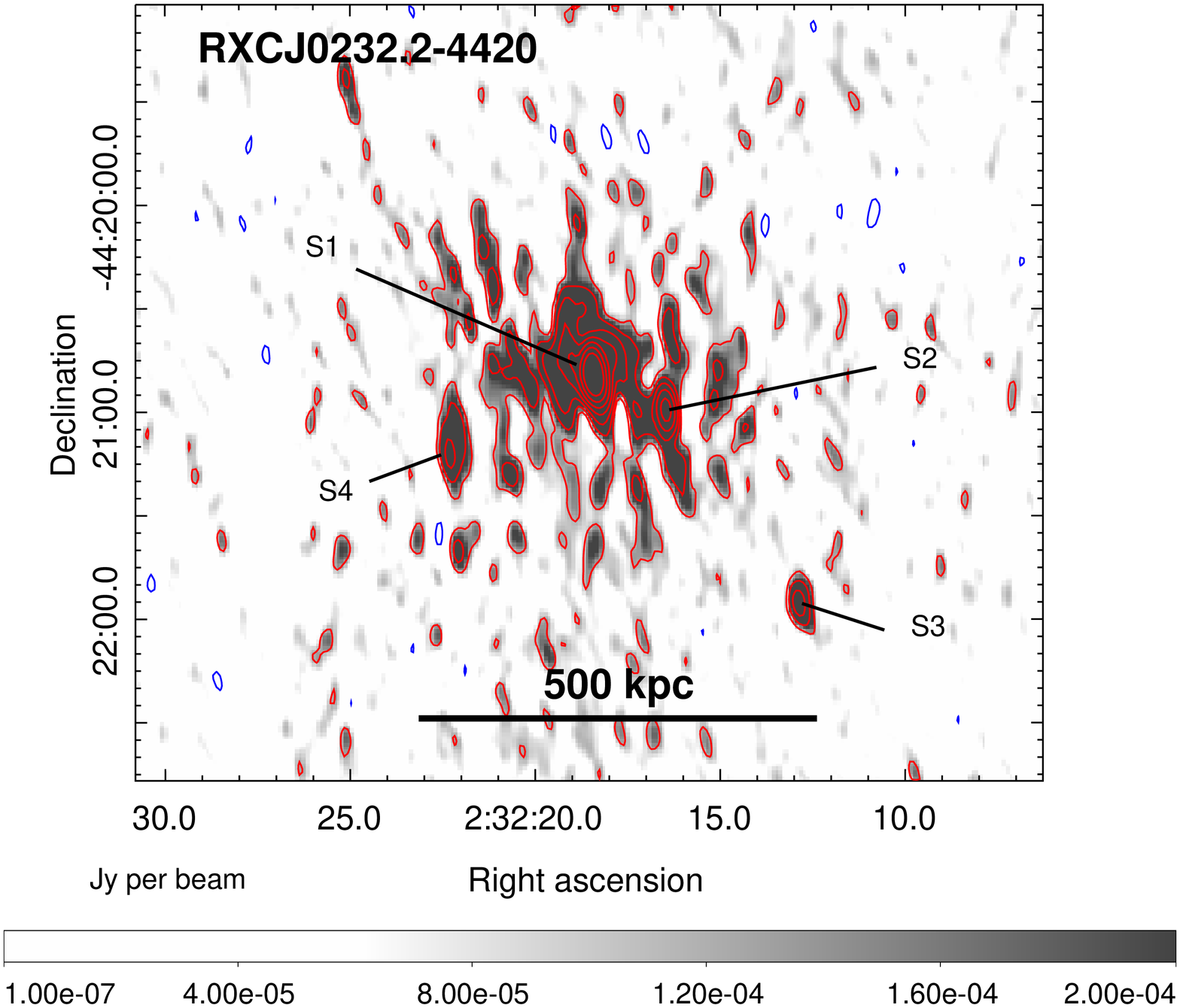}
    \caption{The GMRT 606 MHz image with a resolution of $9.2'' \times 3.9''$, position angle (p. a.) $6.3^{\circ}$ and rms of $40 \mu$ Jy beam$^{-1}$ is shown in grey-scale (0.1 - 500 $\mu$ Jy beam$^{-1}$) and in contours. The red contours are positive and blue are negative. The contours levels are $\pm 0.12, 0.24, 0.48, 0.96, ...$ mJy beam$^{-1}$. The discrete sources S1 - S4 are labelled.}
    \label{rob0}
\end{figure}

\begin{figure}
    \centering
        \includegraphics[width= 8 cm]{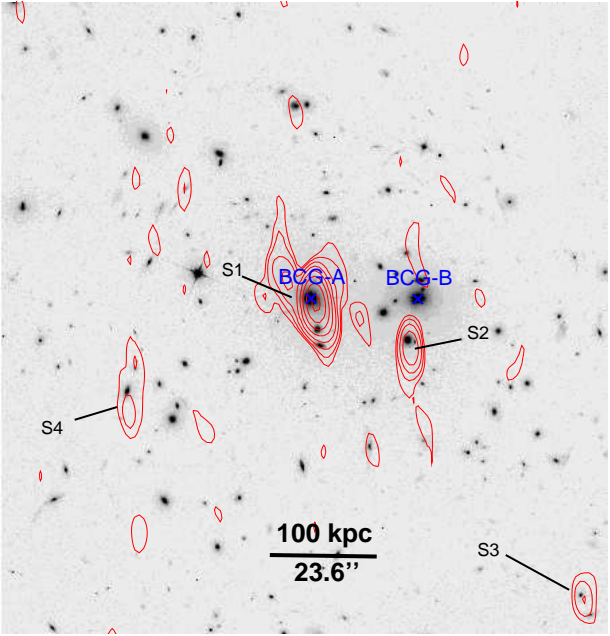}
    \caption{The Hubble Space Telescope WFC3 image {with the filter F160W} is 
    shown in grey-scale with the contours from the GMRT 606 MHz image 
    using baselines longer than $5$k$\lambda$. The contours are at $0.15, 0.3, 0.6, ...$ mJy beam$^{-1}$.
    The beam is $7.7'' \times 3.4''$, position angle (p. a.) $5^{\circ}$.
  The discrete sources S1 - S4 are labelled and the positions of BCGs, A and B are marked with crosses.
  \label{hst}}
\end{figure}

The GMRT 606 MHz image with robust$=0$ weighting of the visibilities is shown in Fig.~\ref{rob0}. It has an rms noise of $0.04$ mJy beam$^{-1}$. The discrete radio sources in the field within two arcminutes 
around the centre are labelled as S1, S2, S3 and S4. The source S1 is associated with the BCG-A in 
the cluster and shows an extension towards northwest; it is unclear at this resolution if 
it is a jet (Fig.~\ref{hst}). The source S4 is marginally resolved and does not show association 
with any galaxy. The sources S2 and S3 are unresolved. The 606 MHz flux densities and the sources within the beam found from the NASA Extragalactic Database (NED) that are the most likely counterparts of these discrete radio sources are given in Table~\ref{radsrc}.

\begin{figure}
    \centering
                \includegraphics[width= 8 cm]{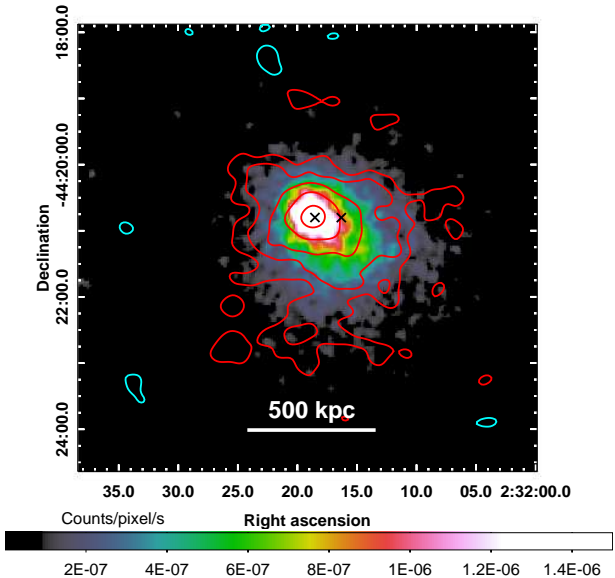}
    \caption{
     The colour scale shows the Chandra (ObsID 4993) 0.5 - 7 keV image with the contours of radio emission 
    at 606 MHz overlaid. The colour scale range is $1\times10^{-9} - 1.5\times10^{-6}$ counts pixel$^{-1}$ s$^{-1}$. The pixel size is $1.97^{\prime\prime}$.
    The GMRT 606 MHz image with a resolution of $20'' \times 20''$ and rms of $0.1 $ mJy beam$^{-1}$. The contour levels are $\pm 0.3, 0.6, 1.2, ...$ mJy beam$^{-1}$. The red contours are positive and cyan are negative. The BCG A and B positions are marked with crosses.}
    \label{sub}
\end{figure}

We have discovered extended diffuse radio emission surrounding the BCG-A in this cluster. 
Based on the location surrounding the BCG at the peak of X-ray emission 
from the ICM it looks like a typical mini-halo. However the largest extent 
of this emission is 800 kpc which is nearly that of giant radio halos 
($\gtrsim 700$ kpc, \citep{cas13}). We refer to this radio emission as a radio halo (RH). 
The RH is roundish at the centre and shows an extension
to the southeast (Fig.~\ref{sub}). 

The RH has a flux density of $52\pm5.0$ mJy at 606 MHz ($S_{606\mathrm{MHz}}$) as measured within 
the area covered by the contour at $3\sigma = 0.3$ mJy beam$^{-1}$ in the low resolution ($20''$) 
image. The error on the flux density was calculated according to $\sqrt{(\sigma \sqrt N_{\mathrm b})^2 + (\sigma_{\mathrm{abs}} S_{606\mathrm{MHz}})^2}$,
where $N_{\mathrm b}$ is the number of beams in the extent of the emission, $\sigma_{\rm abs}$ is the percentage error on
the absolute flux density scale. We used $\sigma_{\rm abs} = 0.1$ \citep[e. g.][]{kal16} and found $N_{\rm b} = 75.6$ to get the 
error of $5$ mJy on the 606 MHz flux density of the RH. The K-corrected radio power at frequency $\nu$ is given by,
$P_{\nu} = (4\pi D_{\rm L}^2) S_{\nu} (1+z)^{(\alpha - 1)}$, where $D_{\rm L}$ is the luminosity distance. 
The 606 MHz radio power of the RH is 
$P_{606\mathrm{MHz}} = 13.9\pm 1.2 \times 10^{24}$ W Hz$^{-1}$. Assuming a spectral index of $1.3$ that is typical 
for radio halos \citep{bru14} we obtain 1.4 GHz radio power of the RH to be, 
$4.6 \times 10^{24}$ W Hz$^{-1}$.

The spectral indices of the diffuse source and that of the BCG  could not be determined definitively.
We found Sydney University Molonglo Sky Survey 843 MHz image \citep[SUMSS,][]
{2003MNRAS.342.1117M}, Murchison Widefield Array (MWA) GaLactic and Extragalactic All-sky MWA  survey 
50 - 200 MHz \citep[GLEAM][]{2015PASA...32...25W} and TIFR GMRT Sky Survey Alternative Data Release 150 MHz \citep[TGSS-ADR][]{2017A&A...598A..78I} images of this region. The TGSS-ADR 147.5 MHz image had a resolution of $57''\times25''$, position angle $0^{\circ}$ and the flux density using JMFIT over the source detected around the centre of the cluster is $295 \pm 10$ mJy. The GLEAM catalogue flux densities of the source J023218-442052 are $228\pm14$ mJy at 223-231 MHz and $1498\pm67$ mJy at 72-80 MHz. These imply a spectral index of $1.7$. The GLEAM beam is $134.7''\times131.6''$, position angle $-30^{-\circ}$ at 200 MHz and thus 
cannot resolve the sources S1-S4. The SUMSS catalogue flux density of the central source at 843 MHz is $50.0\pm3.1$ mJy \citep{2008yCat.8081....0M}. Between SUMSS and TGSS-ADR the spectral index is $1.0$. The steeper spectrum within the GLEAM bands may be contribution from the RH. However images with better resolution are needed to separate the discrete and the diffuse emission to determine the spectral index of the diffuse emission.

\begin{table*}
	\centering
	\caption{Radio sources in the \myclus field.}
	\label{radsrc}
	\begin{tabular}{ccccl} 
		\hline
Label	& RA	& Dec	& $S_{606\mathrm{MHz}}$	& Identification \\
        &       &       &   mJy                 &\\
\hline
S1	& 02h32m18.421s	& -44d20m49.15s	& $28.05 \pm 0.71$ & 2MASS J02321857-4420482 \\	
	&		&		&		   & GALEXASC J023218.4-442047.1 \\
S2	& 02h32m16.471s	& -44d20m59.53s	& $2.67 \pm 0.21$  & 2MASS J0231666-4420570 \\
S3	& 02h32m12.854s	& -44d21m55.31s	& $0.89 \pm 0.08$  & MRSS246-016797 \\
	&		&		&		   & CXOGBAJ023212.9-442154 \\
	&		&		&		   & GALEXASCJ023212.98-442153.9 \\
S4	&02h32m22.235s	& -44d21m09.36s & $1.74 \pm 0.20$  &- \\	
		\hline
	\end{tabular}
\end{table*}

\section{Discussion}
\subsection{Dynamical state of \myclus}
\begin{figure}
    \centering
\includegraphics[width=7cm]{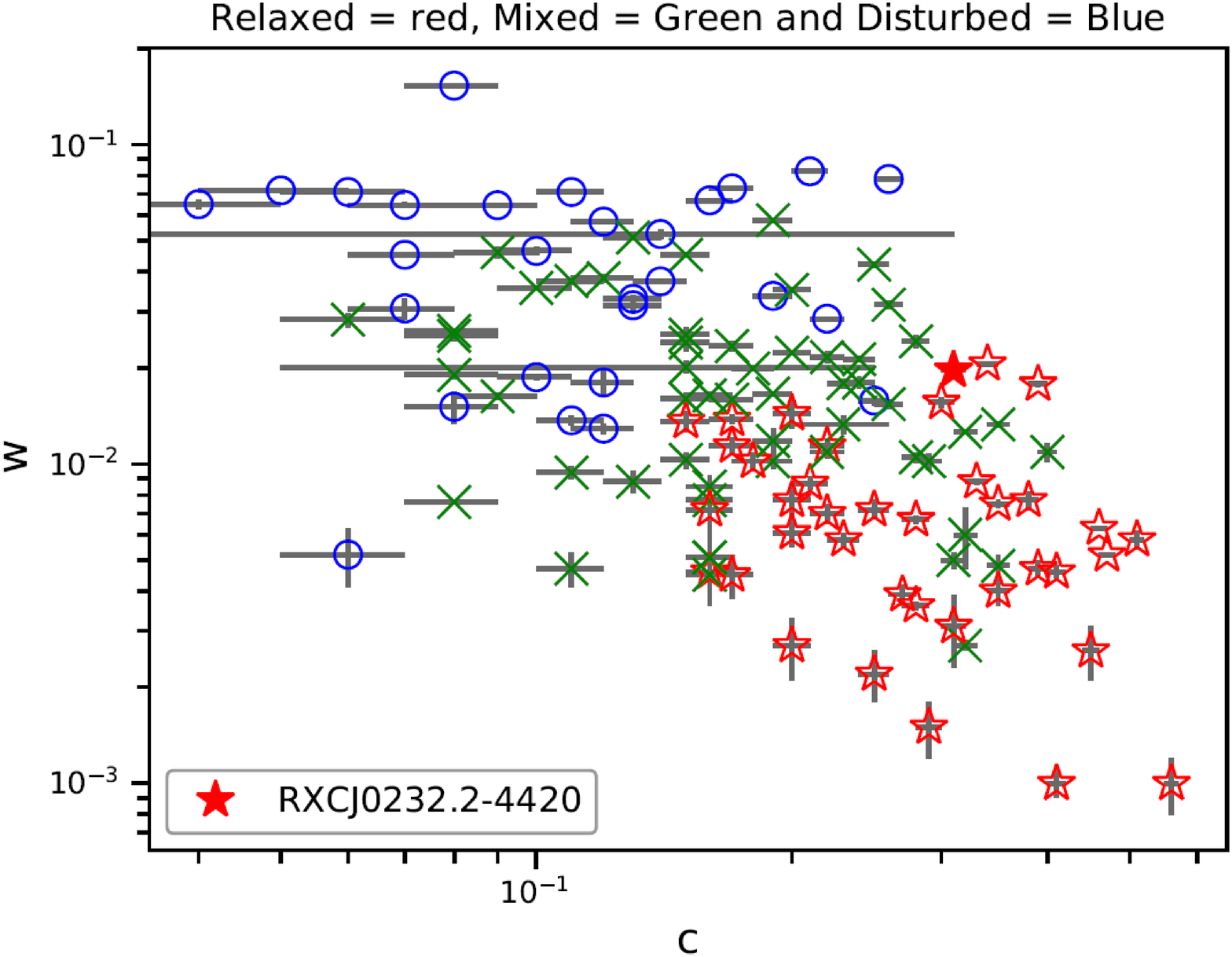}
    \caption{The centroid shift ($w$) versus concentration ($c$) parameter plotted for the sample of clusters presented by \citep{2017ApJ...846...51L}. The cluster \myclus is in their sample and is highlighted with a filled star symbol. It has been classified as a relaxed cluster according to their classification scheme.}
    \label{lovi17}
\end{figure}

In the context of the origin of the diffuse radio emission, the dynamical state of the 
cluster is important. 
{
\citet{2017ApJ...846...51L} have carried out a detailed X-ray morphology analysis of a large sample ($189$) of clusters, including \myclus\, using XMM-Newton images. Among eight different parameters that they calculate to infer the dynamical state of the cluster, they conclude that centroid shift ($w$)  and concentration ($c$) are the best estimators of dynamical state.
The $w-c$ plane is shown in Fig.~\ref{lovi17} for their sample where \myclus\, is classified as a "relaxed" cluster. In the Archive of Chandra Cluster Entropy Profile Tables (ACCEPT \footnote{\url{https://web.pa.msu.edu/astro/MC2/accept/}}) \citep{cav09}, the central temperature of 
\myclus\, is reported to be $5.71\pm0.77$ keV and the central cooling time to be $1.38\pm0.19$ Gyr. This central cooling time is consistent with a "weak" cool-core cluster \citep[e. g.][]{Hudson2010}. According to the classification criteria based on central gas density used in \citet{2011A&A...536A..11P}, this cluster has been classified as a "cool-core" cluster. Thus the properties of the core show evidence of cooling.
\citet{Weissmann2013} have discussed morphological classification of galaxy clusters based on the 
power ratios ($P_3/P_0$) and centroid shifts ($w$) using observations with the XMM Newton for a sample 80 clusters that includes \myclus. This cluster is classified as "complex'' which in their definition is a cluster that does not have two maxima in the X-ray surface brightness but has a complex global structure.  Thus the  works discussed above do not categorise this cluster to be a major merger but do indicate the presence of departure from strong cool-core clusters. The presence of two BCGs in the cluster further indicates a merger in the past.
}

\subsection{Mini-halo to radio halo transition system}
The populations of clusters hosting giant radio halos and mini-halos 
are distinct in their dynamical properties \citep[e. g.][]{cas10,kal15}.
 It has been proposed that a cool-core cluster with a mini-halo may transition into a radio halo cluster 
 if it undergoes a merger \citep{bru14}. While mini-halos could have significant emission due to relativistic electrons that are 
 byproducts of hadronic collisions \citep{2014MNRAS.438..124Z}, a merger can trigger transport of 
 the electrons and turbulent re-acceleration 
 in the outskirts leading to the formation of a radio halo \citep{bru14, wee19}. The morphology of the RH is nearly symmetric about the BCG-A and shows a southern extension that could be 
 following a sloshing cold-front as found in the case of the mini-halo RX J$1720.1+2638$ \citep{2014ApJ...795...73G}.
 
\begin{figure}
    \centering
 \includegraphics[trim={1.0cm 1.0cm 2.5cm 1.0cm},width=7.5cm]{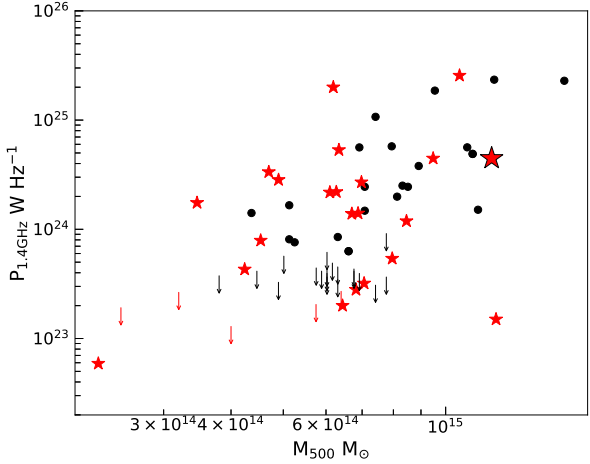}
    \caption{The radio halos (black dots) from \citet{cas13} and mini-halos (small red stars) from \citet{gia14} are plotted in the $P_{1.4\mathrm{GHz}} - M_{500}$ plane. The radio halo upper limits are shown as black arrows and mini-halo upper limits are shown as red arrows \citep{kal15}. The big red star symbol marks \myclus.}
    \label{rhmh}
\end{figure}

 \citet{gia17} have studied the distribution of central entropies ($K_0$) in galaxy clusters with mini-halos and radio halos. They found that the clusters hosting mini-halos have $K_0 < 20$ keV cm$^{2}$ and those hosting radio halos typically have $K_0 > 50$ keV cm$^{2}$. With the $K_0 = 44.62 \pm 12.42$ keV cm$^{2} $, 
 this cluster is in the intermediate zone of radio halos and mini-halos. This is further evidence of 
 it being a transition system. We compared the radio power of RH with that of the mini-halos and radio halos in the $P_{1.4\mathrm{GHz}} - M_{500}$ plane (Fig.~\ref{rhmh}). The RH radio power is well within 
 the typical powers of radio halos as is also the case for a few other mini-halos. It is among the highest mass clusters hosting mini-halos. 
 
 The diffuse radio sources in the clusters Abell 2142, Abell 2390, Abell 2261 and CL$1821+643$  have been classified as peculiar Mpc sized radio sources in cool-core clusters. A two component radio halo is reported in Abell 2142 and is thus unlike typical mini-halos \citep{ven17}. The systems Abell 2390, Abell 2261 \citep{som14} and CL$1821+643$ \citep{bon14,kal16} {are similar to the radio halo reported here. 
 Recently low brightness and steep spectrum emission has been found surrounding the known mini-halos 
 in the clusters PSZ1G139.61+24.20 \citep{Savini18} and RXJ1720.1+2638 \citep{Savini19}. It has been suggested by the authors that sloshing may result in triggering acceleration on large scales. We await our Upgraded GMRT observations to find the spectral index distribution across the radio halo in \myclus and further analysis of the X-ray data to shed more light on the origin of the radio halo.}

\section{Conclusions}
 We report the discovery of diffuse radio emission of size $550\times800$ kpc$^{2}$ surrounding 
    the primary BCG in the cluster \myclus using 606 MHz observations from the GMRT. The 606 MHz flux density of the diffuse radio source is $52\pm5$ mJy.
   The diffuse source on one hand is similar to mini-halos in cool-core clusters that surround a BCG and on the other has an extent typical of radio halos. 
    The dynamical state of the cluster based on the X-ray morphology has been termed as "relaxed'' and 
     complex making it a transition system between a merger and a cool-core.
    The 1.4 GHz radio power of the RH is $4.6 \times 10^{24}$ W Hz$^{-1}$ assuming a spectral index of 1.3. In the $P_{1.4\mathrm{GHz}} - M_{500}$ plane, it falls within the general trend of the radio halos. Mini-halos are more scattered in this plane and if classified as a mini-halo, it will be the second most massive cluster to host it. The RH properties support the scenario that it is rare system where a mini-halo is transitioning into a radio halo. 
\section*{Acknowledgements}
We thank the referee for their comments that improved the clarity of the paper.
RK acknowledges support through the DST-INSPIRE 
Faculty Award by the Government of India.
We thank the staff of the GMRT that made these observations possible. 
GMRT is run by the National Centre for Radio Astrophysics of the Tata Institute 
of Fundamental Research. Based on observations made with the NASA/ESA Hubble Space Telescope, and obtained from the Hubble Legacy Archive, which is a collaboration between the Space Telescope Science Institute (STScI/NASA), the Space Telescope European Coordinating Facility (ST-ECF/ESA) and the Canadian Astronomy Data Centre (CADC/NRC/CSA). This research has made use of 
the NASA/IPAC Extragalactic Database (NED) which is operated by the Jet 
Propulsion Laboratory, California Institute of Technology, under contract with 
the National Aeronautics and Space Administration. This research made use of 
Astropy, a community-developed core Python package for Astronomy (Astropy 
Collaboration, 2018). The scientific results reported in this article are based 
in part on data obtained from the Chandra Data Archive.





\bibliographystyle{mnras}
\bibliography{ruta_all_1} 




\bsp	
\label{lastpage}
\end{document}